\documentstyle[graphicx,twocolumn,prl,aps]{revtex}
\tightenlines
\begin{document}
\draft
\title{Dissociation of vertical semiconductor diatomic
artificial molecules }
\author{ M. Pi$^1$, A. Emperador$^1$,
M. Barranco$^1$\thanks{corresponding author: manuel@ecm.ub.es},
 F. Garcias$^2$, K. Muraki$^3$, S.
Tarucha$^{3,4}$,
D. G. Austing$^3$\thanks{corresponding author:
austing@will.brl.ntt.co.jp}
}
\address{$^1$Departament ECM,
Facultat de F\'{\i}sica, Universitat de Barcelona, E-08028 Barcelona,
Spain}
\address{$^2$Departament de F\'{\i}sica, Facultat de
Ci\`encies, Universitat de les Illes Balears, E-07071 Palma de
Mallorca, Spain}
\address{$^3$NTT Basic Research Laboratories, 3-1
Morinosato Wakamiya, Atsugi, Kanagawa, 243-0198, Japan} %
\address{$^4$Departament of Physics and ERATO Mesoscopic Correlation
Project, University of Tokyo, 7-3-1 Hongo, Bunkyo-ku, Tokyo, 113-0033,
Japan}
\date{\today}

\maketitle

\begin{abstract} We investigate the dissociation of few-electron
circular vertical semiconductor double quantum dot artificial
molecules at 0 T as a function of interdot distance.
Slight mismatch introduced in the fabrication of the
artificial molecules from nominally identical constituent quantum wells
induces localization by offsetting the energy levels in the quantum dots by up
to 2 meV,  and this plays a crucial role in the appearance of the addition
energy spectra as a function of coupling strength particularly in the
weak coupling limit.

\narrowtext

\end{abstract}

Semiconductor quantum dots (QD's) are widely considered as artificial
atoms, and are uniquely suited to study fundamental electron-electron
interactions and quantum effects\cite{GEN}. There are many analogies
with `natural' atoms. One of the most appealing is the capability of
forming molecules. Indeed, systems composed of two QD's, `artificial'
quantum molecules (QM's), coupled either laterally or vertically, have
recently been investigated experimentally\cite{latexp,SAD} and
theoretically\cite{lateor,vertical,Par00,TOK}. Nevertheless, the direct
observation of a systematic change in the addition energy spectra for
few-electron (number of electrons, $N<13$) QM's as a function of
interdot coupling, has not been reported, and calculations of QM
properties widely assume a priori that the constituent QD's are
identical\cite{lateor,vertical,Par00}. Special transistors incorporating
QM's\cite{Aus98} made by vertically coupling two well defined and highly
symmetry QD's\cite{Tar96} are ideally suited to observe the former and
test the latter.

In this work we present experimental and theoretical addition energy
spectra characterizing the dissociation of slightly asymmetric vertical
diatomic QM's on going from the strong to the weak coupling
limits that correspond to small and large interdot distances, $b$,
respectively.  We also show that spectra calculated for symmetric diatomic
QM's only resemble those actually observed when the coupling is
strong. The interpretation of our experimental results is based on the
application of  local-spin density-functional theory (LSDFT)
\cite{Per81,DFT/SD,Pi01}. It follows the development
of the method thoroughly described in Ref.  \cite{Pi01}, which
includes finite thickness effects of the dots, and uses a relaxation
method to solve the partial differential equations arising from a
high order discretization of the Kohn-Sham equations on a spatial mesh
in cylindrical coordinates\cite{SIC}. Axial symmetry  is
imposed, and the exchange-correlation energy has been taken from
Perdew and Zunger\cite{Per81}.

The molecules we study are formed by coupling, quantum
mechanically and electrostatically, two QD's
which individually can show clear atomic-like
features \cite{Aus98,Tar96}.  For the materials we
typically use, the energy splitting between the bonding and
anti-bonding sets of single particle (sp) molecular states,
$\Delta_{\rm SAS}$, can be
varied from about 3.5 meV for $b=2.5$ nm (strong coupling) to about
0.1 meV for $b=7.5$ nm (weak coupling)\cite{Aus98}.  This is expected
to have a dramatic effect on the electronic properties of QM's
\cite{vertical,Par00,TOK,Pi01}.
Figure 1 shows (a) a schematic diagram of a
sub-micron circular mesa, diameter $D$, containing two vertically
coupled QD's, and (b) a scanning electron micrograph of a typical mesa
after gate metal deposition.  The starting material, a special triple
barrier resonant tunneling structure, and the processing recipe are
described elsewhere\cite{Aus98,SST}.  Current $I_{\rm d}$ flows
through the  two QD's, separated by the central barrier of
thickness $b$, between
the substrate contact and grounded top contact in response to voltage
$V_{\rm d}$ applied to the substrate, and gate voltage $V_{\rm g}$.  The
structures are cooled to about 100 mK and no magnetic field is
applied.

\begin{figure}
  \centering
     \includegraphics[totalheight=100mm,width=75mm]{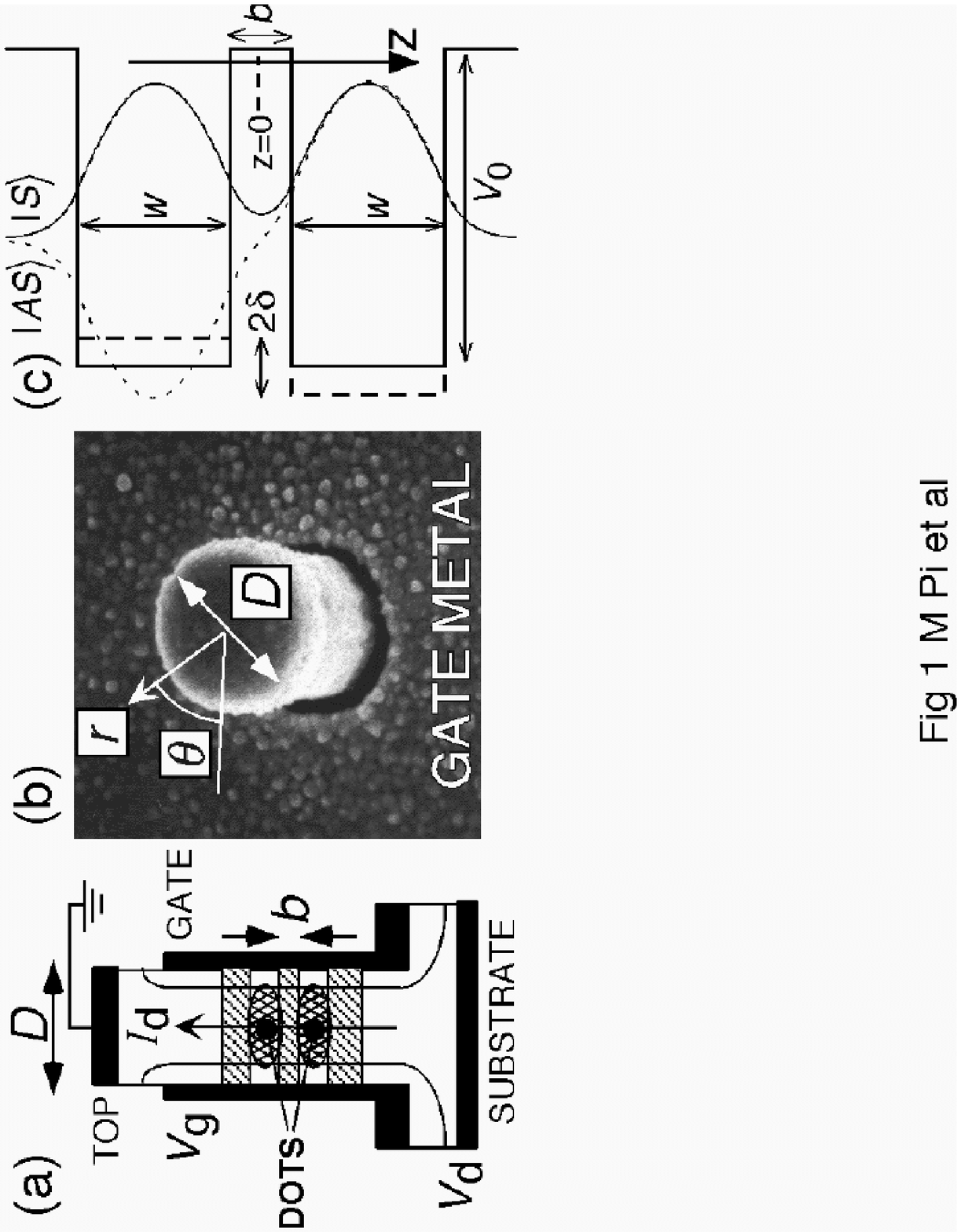}
\vspace{1cm}
\caption[]{ Schematic diagrams of (a) mesa containing
two vertically coupled quantum dots
and (c) double quantum well structure, and
(b) scanning electron micrograph of a typical circular mesa.
 }
\label{fig1}
\end{figure}

To analyze the experiments we have modeled the QM by two axially
symmetric QD's.  The QM is confined in the radial direction
by a harmonic oscillator potential $m \omega^2 r^2/2$ of
strength $\hbar \omega=5$ meV (a realistic lateral
confinement energy for a single QD in the few electron
limit\cite{Tar96,nota}), and in
the axial ($z$) direction by a double quantum well structure whose
 wells are of same width $w$, and have depths $V_0 \pm \delta$, with
$\delta \ll V_0$\cite{material}.  Figure \ref{fig1} (c)
schematically shows the double quantum well structure and its unperturbed
bonding and anti-bonding sp wavefunctions.
We have taken $V_0=225$ meV and $w=12$ nm, which are appropriate for
the actual experimental devices.  If $\delta$ is set to zero, the
artificial molecule is symmetric (`homonuclear' diatomic QM);
otherwise, it is asymmetric (`heteronuclear' diatomic QM).  In the
calculations here, $\delta$ is 0, or is set to a realistic value
of  0.5 or 1 meV \cite{UPEXP}.  In the
homonuclear case $\Delta_{\rm SAS}$ is well reproduced by the law
$\Delta_{\rm SAS}(b) = \Delta_0 \, e^{-b/b_0}$ with $b_0=1.68$ nm, and
$\Delta_0=19.1$ meV.  It is easy to check that in
the weak coupling limit $2 \delta$ is approximately the
energy splitting between the bonding and anti-bonding sp states which
would be  almost degenerate if $\delta$ is 0.  For this reason we call
the mismatch (offset) the quantity $2 \delta$.

\begin{figure}
\centering
\includegraphics[angle=270,totalheight=55mm,width=150mm]{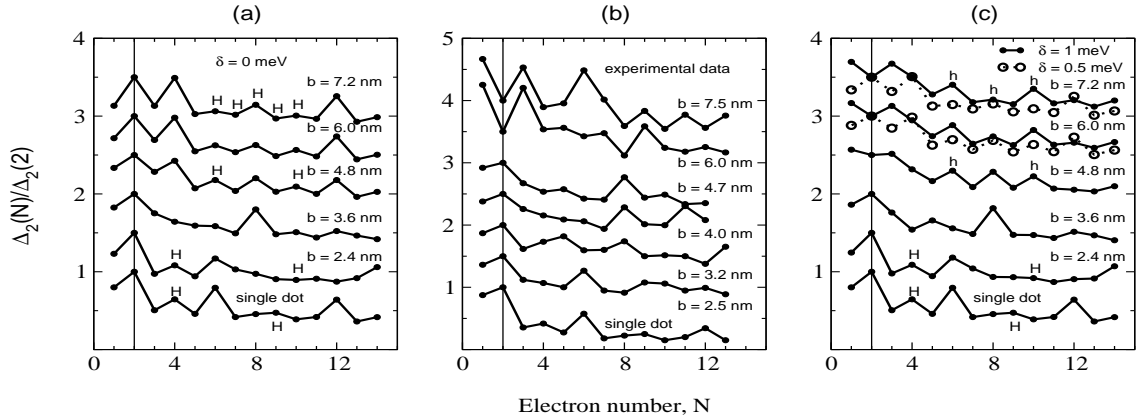}
\vspace{1cm}
\caption[]{ (a) Calculated
$\Delta_2(N)$/$\Delta_2(2)$ for homonuclear QM's with
different interdot distances, $b$.
Also shown is the calculated
reference spectrum for a single QD.
(b) Experimental QM addition energy spectra,
$\Delta_2(N)$/$\Delta_2(2)$, for several interdot distances between 2.5
and 7.5 nm. Also shown is an experimental reference
spectrum for a single QD\cite{MAT}.
(c) Same as panel (a) but for heteronuclear QM's
obtained using a $2 \delta=2$ meV mismatch
(dotted lines for $b=6.0$ and 7.2 nm are for $2 \delta=1$ meV).
In each panel the curves have been vertically offset so that at $N=2$
they are equally separated by 0.5 units for clarity.
All traces in panels (a) and (c) except 3.6 and 6.0 nm:
$H (h)$ marks cases where we could clearly identify
Hund's first rule like filling within
single dot, or bonding or anti-bonding states (constituent dot
states).
}
\label{fig2}
\end{figure}

Figure \ref{fig2}(a) shows calculated addition energy spectra,
$\Delta_2(N) = U(N+1) -2 U(N) + U(N-1)$, for homonuclear QM's with
realistic values of $b$ conveniently
normalized as $\Delta_2(N)/\Delta_2(2)$.
$U(N)$ is the total energy of the $N$-electron system.
$\Delta_2(N)$ can reveal a
wealth of information about the energy required to place an extra
electron into a QD or QM system \cite{Tar96,MAT}.  For small $b$
($\Delta_{\rm SAS} \gtrsim \hbar \omega$)
the spectrum of a few electron QM is rather similar to a
single QD, at least for $N<7$\cite{nota}.
At intermediate dot
separation, the spectral pattern  becomes more complex.
However, a simple picture emerges at larger interdot distances when
the molecule is about to dissociate.  For example, at $b=7.2$ nm
strong peaks at $N=2$, 4, 12, and a weaker peak at $N=8$ appear that
can be easily interpreted from the peaks appearing in the single QD
spectrum.  The peaks at $N=4$ and 12 in the QM are a consequence of
symmetric
dissociation into two closed shell (magic) $N=2$ and 6 QD's
respectively, whereas the peak at $N=8$ corresponds to the
dissociation of the QM into two identical stable QD's
holding four electrons each filled according to Hund's first rule
to give maximal spin \cite{Par00,Tar96}.
The QM peak at $N=2$ is related to the localization of one
electron on each constituent dot, the two-electron state being a
spin-singlet QM configuration.

Since the modeled QM is homonuclear, each sp wavefunction  is shared
$50\%-50\%$ between the
two constituent QD's.  Electrons are completely delocalized in
the strong  coupling limit.  As $b$ increases, $\Delta_{\rm SAS}$
decreases and eventually bonding, $|{\rm S}\rangle$, and anti-bonding,
$|{\rm AS}\rangle$,
sp molecular states become quasi-degenerate.  Electron localization
can thus be achieved combining these states as $(|{\rm S}\rangle \pm
|{\rm AS}\rangle)/\sqrt{2}$.

We conclude from Fig.  \ref{fig2}(a) that the fingerprint of a
dissociating few-electron homonuclear diatomic QM is the appearance of
peaks in $\Delta_2(N)$ at $N=2$, 4, 8 and 12\cite{Par00}.
This is a robust
statement, as it stems from the well understood shell structure of a
single QD.  If we now compare this picture with the experimental
spectra shown in Fig.  \ref{fig2}(b), we are led to conclude that the
experimental devices are not homonuclear, but heteronuclear QM's.

The origin of the mismatch is the difficulty in fabricating two
perfectly identical constituent QD's in the QM's discussed here, even
though all the starting materials incorporate two nominally identical
quantum wells.  This mismatch can clearly influence the degree of
delocalization-localization, and the consequences will depend on how
big $2 \delta$ is in relation to $\Delta_{\rm SAS}$ \cite{TOK,MUR}.
Elsewhere we will discuss how the effective value of
$\Delta_{\rm SAS}$ is measured, and the mismatch is determined for all
values of $b$\cite{UPEXP}, so we merely note here that $2 \delta$
is typically 0.5
to 2 meV and nearly always with the upper QD (nearest top contact of
mesa) states at higher energy than the
corresponding lower QD states [see Fig. \ref{fig1}(c)].

Figure \ref{fig2}(b) shows experimental spectra, also
normalized as $\Delta_2(N)/\Delta_2(2)$, for QM's with $b$ between
2.5 and 7.5 nm, deduced accurately from peak spacings between Coulomb
oscillations ($I_{\rm d}-V_{\rm g}$) measured by applying
an arbitrarily small bias ($V_{\rm d}<100\, \mu$V).
Likewise, also shown is a reference spectrum for a single QD
\cite{MAT}.  The diameters of the mesas lie in the range of
0.5 to 0.6 $\mu$m, and while all mesas are circular, we can not exclude
the possibility that the QM's and QD's inside the mesas may
actually be slightly
non-circular, and that the confining potential is not perfectly parabolic
as $N$ increases \cite{MAT,Aus99}.  We emphasize the following:  i)
The spectrum for the most strongly coupled QM ($b=2.5$ nm) resembles
that of the QD up to the third shell ($N=12$). ii) For intermediate
coupling ($b=3.2$ to 4.7 nm), the QM spectra are quite different from
the QD spectrum, and a fairly noticeable peak appears at $N=8$. iii)
For weaker coupling ($b=6.0$ and 7.5 nm) the spectra are different
again, with prominent peaks at $N=1$ and 3.

We confirmed the heteronuclear character of the QM's by performing LSDFT
calculations with a 2 meV mismatch.  The results are
displayed in Fig.  \ref{fig2}(c).  For $b=6.0$ and 7.2 nm, spectra
for a 1 meV mismatch are also given.  One-to-one comparison
between theory and experiment of {\em absolute} values is not helpful,
because the QM's (QD's) actually
behave in a very complex way\cite{nota}.
In particular, $2 \delta$ can vary from
device-to-device, and probably it decreases with $N$\cite{UPEXP}.
Nonetheless, the overall agreement between theory and experiment
of the general spectral shape  is
quite good, indicating the crucial role played by mismatch.  In
particular, the appearance of the spectra in the weak coupling limit
for small $N$ values is now correctly given, as well as the evolution with
$b$ of the peak appearing at $N=8$ for intermediate coupling.  A
comparison between panels (a) and (c) of
Fig.  \ref{fig2} reveals that for
smaller values of $b$ ($\lesssim 4.8$ nm), for a reasonable choice of
parameters $(\omega,\delta)$, mismatch does not produce sizeable
effects.  The reason is
that the electrons are still rather delocalized, and distributed
fairly evenly between the two dots.  Exceptions to this substantial
delocalization may arise
only when both the constituent single QD states are magic, as discussed
below, at intermediate coupling.
For larger interdot distances, mismatch induces electron localization.
The manner
in which it happens is determined by the balance between interdot and
intradot Coulomb repulsion, and by the degree of mismatch between
the sp energy levels,  and so is difficult
to predict except in some trivial cases for certain model parameters
$(\omega, \delta)$.  For example, a large mismatch compared to $\hbar
\omega$ will cause the QD of depth $V_0-\delta$ to eventually `go away
empty'.

\begin{figure}
\centering
\includegraphics[angle=270,totalheight=55mm,width=150mm]{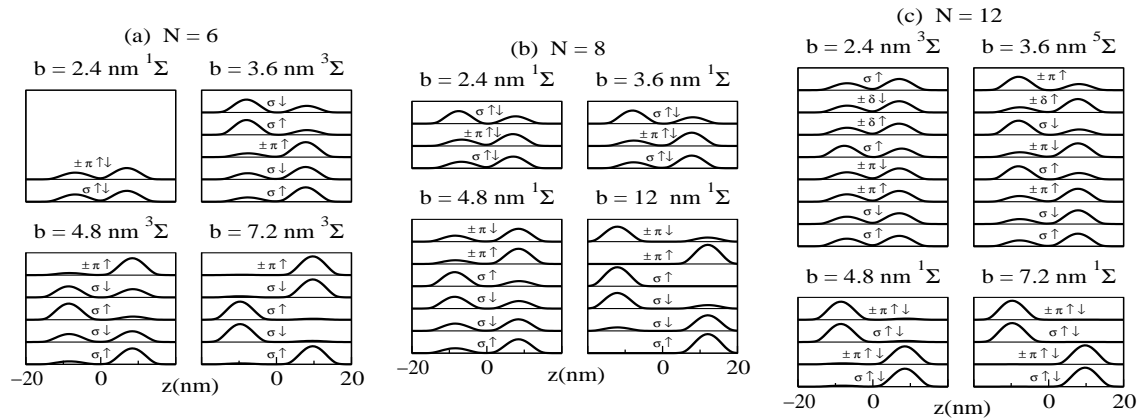}
\vspace{1.5cm}
\caption[]{ Calculated probability
distribution functions ${\cal P}(z)$ (arbitrary units) as a function
of $z$ for the heteronuclear $N=6$, $N=8$, and $N=12$ QM's (a), (b), and
(c) respectively, using a $2 \delta=2$ meV mismatch.  }
\label{fig3}
\end{figure}

Finally, still assuming perfect coherency, a deeper theoretical
understanding of heteronuclear QM
dissociation can now be
gained from analysis of the evolution with $b$ of the sp molecular
wavefunctions.  Thus, for each sp wavefunction
$\phi_{nl\sigma}(r,z,\theta)=u_{nl\sigma}(r,z) e^{-i l \theta}
\chi_{\sigma}$ we introduce a $z$-probability distribution
function defined as

\begin{equation}
{\cal P}(z) \equiv 2 \pi \int_0^{\infty} dr \, r
\,[u(r,z)]^2 \,\,\, .
\label{eq1}
\end{equation}
Figure \ref{fig3}
shows ${\cal P}(z)$ for (a) $N=6$, (b) $N=8$, and (c) $N=12$ (deeper
well always in the $z>0$ region), each for several values of $b$.  States
are labeled as $\sigma, \pm\pi, \pm\delta, \ldots$ depending on the
$l=0,\pm 1, \pm 2, \ldots$ sp angular momentum, and
$\uparrow,\downarrow$ indicate the spins.  In each sub-panel, the
probability functions are plotted, ordered from bottom to top,
according to the increasing energies of the orbitals.  For each $b$,
the third component of the total spin and total orbital angular
momentum of the ground state are also indicated by the standard
spectroscopic notation $^{2 S_z+1}|L_z|$ with $\Sigma, \Pi, \Delta,
\ldots$ denoting $|L_z|= 0, 1, 2, \ldots$ We conclude that:  i) QM's
dissociate more easily for smaller values of $b$, if they yield magic
number QD's, as is the case for $N = 12 \rightarrow 6+6$ for $b=4.8$ nm
(c) or $N = 4  \rightarrow 2+2$ (not shown), for example.
 ii) Particularly for intermediate values of $b$, not
all orbitals contribute equally to the QM bonding, i.e., the degree
of hybridization is not the same for all QD sp orbitals.  See
for example the $\pi$ and $\sigma$ states in the $b=4.8$ nm panel of
(a). iii)
At larger $b$, dissociation can lead to Hund's first rule
like filling in one of the QD's and full shell filling in the
other dot.
See for example the
$b=7.2$ nm panel in (a) for $N=6$, which dissociates into $2+4$.
The same happens for the $N=10$ QM, which dissociates into $4+6$
(not shown).
In other cases, dissociation leads to
Hund's first rule
like filling in each of the QD's, as shown in
the $b=12$ nm panel of (b) for $N=8$, which breaks into $4+4$.
In close
analogy with natural molecules, atomic nuclei,
or multiply charged simple metal clusters\cite{Nah97},
homo and heteronuclear QM's
choose preferred dissociation channels yielding the most stable QD
configurations.
iv) Some
configurations are extremely difficult to disentangle:  even at very
large $b$, there can still be orbitals contributing to the QM
bonding.  A good example of this is the $N=8$ QM for $b=12$ nm (b).

This work has been performed under grants PB98-1247 and PB98-0124 from
DGESIC, and 2000SGR-00024 from Generalitat of Catalunya, and partly
funded by NEDO program (NTDP-98).  We are grateful for the assistance
of T. Honda with processing the samples, and for useful discussions with
K. Ono and S. Sasaki.

\onecolumn
\vspace{2.5cm}

\begin{references}
\bibitem{SIC} Self-interaction corrections\cite{Per81} have not
been included. We have checked\cite{Pi01} that they do not play
an important role in the calculated addition spectra, see Fig. 12 of
this reference.
\bibitem{material} We have taken for the dielectric constant
and  the electron effective mass values
corresponding to GaAs, i.e., $\epsilon=12.4$
and $m^*=0.067$.
\end{references}
\end{document}